\documentclass[final]{cvpr}

\usepackage{times}
\usepackage{epsfig}
\usepackage{graphicx}
\usepackage{amsmath}
\usepackage{amssymb}
\usepackage{multirow}

\usepackage[pagebackref=true,breaklinks=true,colorlinks,bookmarks=false]{hyperref}

\def\cvprPaperID{2664} 
\def\confYear{CVPR 2022}

\begin{document}
	
	\title{Learning the Degradation Distribution for Blind Image Super-Resolution}
	
	\author{
		Zhengxiong Luo\textsuperscript{1,2,3}\hspace{0.5em}
		\thanks{Yan Huang is the corresponding author} Yan Huang\textsuperscript{2,3}\hspace{0.5em}
		Shang Li\textsuperscript{1, 3}\hspace{0.5em}
		Liang Wang\textsuperscript{2,3}\hspace{0.5em}
		Tieniu Tan\textsuperscript{2,3}\hspace{0.5em}\\
		\textsuperscript{1}University of Chinese Academy of Sciences (UCAS)\\
		\textsuperscript{2} National Laboratory of Pattern Recognition (NLPR)\\
		Center for Research on Intelligent Perception and Computing (CRIPAC)\\
		\textsuperscript{3}Institute of Automation, Chinese Academy of Sciences (CASIA)\\
		{\tt\small \quad zhengxiong.luo@cripac.ia.ac.cn\quad \{yhuang, wangliang, tnt\}@nlpr.ia.ac.cn \quad lishang2018@ia.ac.cn}}
	
	\maketitle
	
	\begin{abstract}
		
		Synthetic high-resolution (HR) \& low-resolution (LR) pairs are widely used in existing super-resolution (SR) methods.  To avoid the domain gap between synthetic and test images,  most previous methods try to adaptively learn the synthesizing (degrading)  process via a deterministic model. However, some degradations in real scenarios are stochastic and cannot be determined by the content of the image. These deterministic models may fail to model the random factors and content-independent parts of degradations, which will limit the performance of the following SR models. In this paper, we propose a probabilistic degradation model (PDM), which studies the degradation $\mathbf{D}$ as a random variable, and learns its distribution by modeling the mapping from a priori random variable $\mathbf{z}$ to $\mathbf{D}$. Compared with previous deterministic degradation models,  PDM could model more diverse degradations and generate HR-LR pairs that may better cover the various degradations of test images, and thus prevent the SR model from over-fitting to specific ones. Extensive experiments have demonstrated that our degradation model can help the SR model achieve better performance on different datasets. The source codes are released at \url{git@github.com:greatlog/UnpairedSR.git}.
		
	\end{abstract}
	
	\section{Introduction}
	
	Image super-resolution (SR) aims to restore the high-resolution (HR) image from its low-resolution (LR) version. Recently, learning-based SR methods~\cite{srcnn,fsrcnn,lin2022flow} have achieved remarkable performance. These methods usually need paired HR-LR samples as the training dataset. However, it is difficult to acquire such paired samples in real scenarios. As an alternative, synthetic HR-LR pairs are widely used in most existing SR methods. Some of them synthesize data by degrading HR images with predefined settings, such as bicubic downsampling~\cite{li2022dfan,rcan,li2021approaching,luo2021efficient}, convolving with provided blur kernels, or adding noises in a certain range~\cite{ikc,usr,dan,danv2}. But in blind SR, where the degradations of test images are unknown, the predefined degradation settings may likely be different from that of test images. This gap will largely destroy the performance of these methods~\cite{real-esrgan,bsrgan} in real scenarios. 
	
	\begin{figure}[t]
		\centering
		\includegraphics[width=\linewidth]{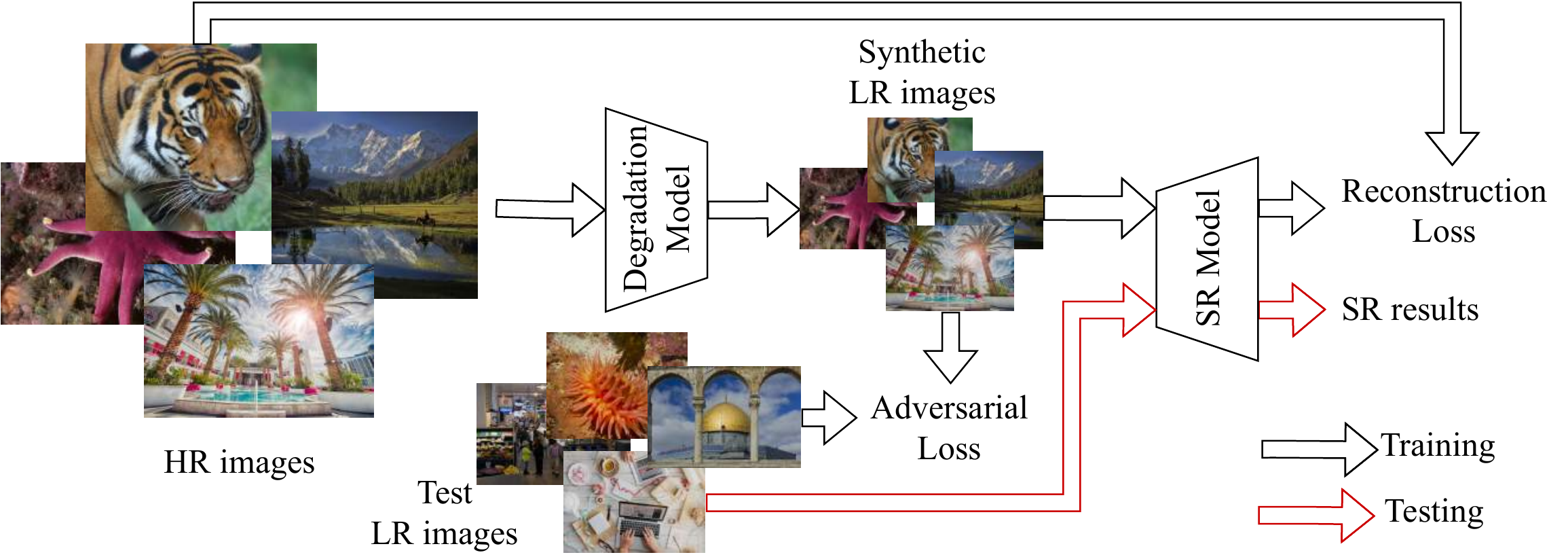}
		\caption{Degradation-learning-based methods usually use adversarial training to encourage a degradation model to produce synthetic LR images in the same domain with  the test images.}\label{learn_deg}
		\vspace{-0.05\linewidth}
	\end{figure}
	
	To avoid the domain gap, some works~\cite{cingan,pseudosr,dasr,cresgan,gan_first} try to adaptively learn the degradation settings. As shown in Figure~\ref{learn_deg}, the degradation model is supervised via adversarial training~\cite{gan} and encouraged to produce LR images in the same domain as the test images. In KernelGAN~\cite{kernel_gan}, the degradation model is designed as linear convolutional layers. And in CycleSR~\cite{cyclesr}, a nonlinear degradation model is learned in a CycleGAN~\cite{cyclegan} framework. These learned degradation models could produce HR-LR pairs that have a smaller gap with the test cases, and thus could be used to train a better SR model.
	
	However, most previous degradation-learning-based SR methods have a common drawback: their degradation models are deterministic, and each HR image can only be degraded to a certain LR image. It implies an assumption: the degradation is completely dependent on the content of the image. However, this may not hold in most cases. Some degradations are content-independent and stochastic, such as random noises or blur caused by random shakes of cameras. These random factors and content-independent parts of degradations could not be well modeled by these deterministic models. \textit{A better assumption is that the degradation is subject to a distribution, which may be better modeled by a probabilistic model}.
	
	Based on above discussions, we propose a probabilistic degradation model (PDM) that could learn the degradation distribution for blind image super-resolution. Specifically, we parameterize the degradation with two random variables, \ie, the blur kernel $\mathbf{k}$ and random noise $\mathbf{n}$, by formulating the degradation process as  a linear function: 
	\begin{equation}
		\mathbf{D}(\mathbf{x}) = (\mathbf{x}\otimes \mathbf{k})\downarrow_{s} + \mathbf{n} \label{downsample},
	\end{equation}
	where $\mathbf{x}$ denotes the HR image, $\otimes$ denotes the convolving operation, and $\downarrow_{s}$ denotes downsampling with scale factor $s$~\cite{usr,dan}. Thus, the distribution of $\mathbf{D}$ can be represented as the joint distribution of  $\mathbf{k}$ and $\mathbf{n}$, which can be modeled by learning the mapping from a priori random variable $\mathbf{z}$ to $\mathbf{k}$ and $\mathbf{n}$. PDM is then trained in a adversarial framework, and the distribution of $\mathbf{D}$ could be automatically learned during the training. In this way, PDM could model the random factors in degradations and better decouple the degradations with the image content, which may be a more practical approximation in real scenarios. As a result, it may be easier for PDM to cover the diverse degradations of all test images and prevent the SR model from over-fitting specific ones. PDM can  serve as a data generator and be easily integrated with existing SR models to help them improve performance in applications.
	
	Our contributions can be summarized as below:
	\begin{itemize}
		\item[1.] To the best of our knowledge, we are the first to study the degradation in blind SR as a random variable and try to learn its distribution via a probabilistic model. It allows us to model more diverse degradations and synthesize training samples closer to the domain of test images.
		\item[2.] We propose a probabilistic degradation model (PDM) that could model the random factors in degradations and better decouple the degradation with image content, which may make it easier to learn the degradation distribution for blind SR.
		\item [3.] We carefully re-implement and study different degradation-learning-based SR methods and provide comprehensive comparisons. The related codes will be released for further research.
		\item[4.] Based on the probabilistic degradation model, we further propose a unified framework for blind SR, which achieves the state-of-the-art performance on mainstream benchmarks datasets.
	\end{itemize}
	
	\section{Related Works}
	
	
	\subsection{Predefined-Degradation-Based}
	Early SR methods typically synthesize HR-LR samples with predefined degradations. The most widely used setting is bicubic downsampling~\cite{rdn,rcan,cai2021mask,deng2022rformer}. However, some researchers begin to realize that this kind of synthetic data has a domain gap with real test images. And this gap will lead to a dramatic performance drop when these methods are applied to real applications~\cite{ikc}. Thus, in~\cite{srmd,dpsr,usr,ikc,dan}, researchers start to synthesize samples with more complicated settings, including multiple blur kernels (isotropic \& anisotropic) and random noises (additive white gaussian noises) with different levels. Larger degradation space grants these models better generalization abilities. But compared with the huge degradation space in real scenarios, the variety of predefined degradations is still limited and these methods still fail in most applications~\cite{blind-survey,bsrgan,cai2021learning}. 
	
	Recently, \cite{real-esrgan} and \cite{bsrgan} further augment the predefined settings to repeated degradations, which largely broadens their application scenarios. However, as a sacrifice for better generalization ability, they may also tend to produce over smooth results, and we will further discuss it in the experiments section (Sec~\ref{compare}).
	
	\begin{figure*}[t]
		\centering
		\includegraphics[width=\linewidth]{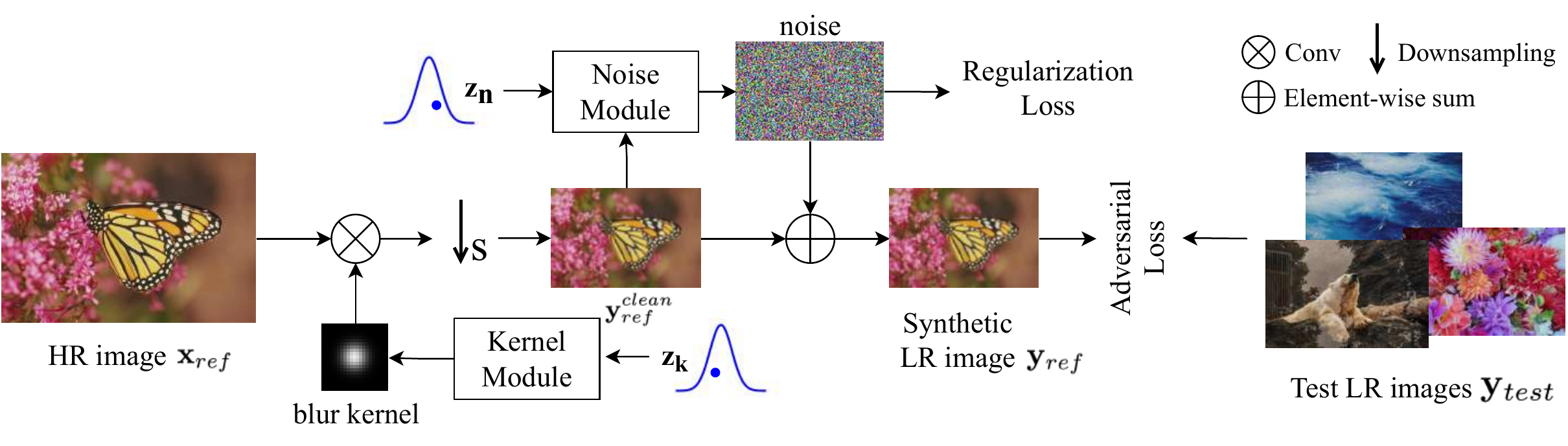}
		\caption{The details of our probabilistic degradation model (PDM)). The discriminator is left out for more intuitive description.}\label{deg_model}
		\vspace{-0.02\linewidth}
	\end{figure*}
	
	\subsection{Degradation-Learning-Based}
	\noindent\textbf{Deterministic degradation model.}
	To generate LR images closer to the domain of real test images,  some methods choose to learn the degradations adaptively. In~\cite{kernel_gan} and~\cite{onsr}, a neural network with linear convolutional layers is used to model the blurring process. This network is trained via adversarial loss~\cite{gan}, which encourages its output to be more like the test image. In this way, the blur kernel of the test image can be learned by this network. In~\cite{fssr}, a more complex (nonlinear) neural network is further used to model the whole degrading process. To get the content better preserved in degraded images,  Chen \etal~\cite{cyclesr} further train the nonlinear neural network model in a CycleGAN~\cite{cyclegan} framework. As their degradations are adaptively learned, these methods can customize an SR model for different sets of test images and usually perform better than predefined-degradation-based methods in real scenarios. However, their degradation models are deterministic. And they may fail to model the random factors in degradations, which potentially limits the performance of their SR models.
	
	\vspace{0.02\linewidth}
	\noindent\textbf{Degradation pool.}
	In~\cite{ji2020real}, Ji \etal model the degradation as a  pool that contains the blur kernels and noises learned from the test images. The degradation pool is then used to synthesize training samples to train the SR model. This method shares a similar idea with our PDM and achieves great performance on the NTIRE2020 challenge. However, the blur kernels and noises in~\cite{ji2020real} are learned respectively, which involves two different tasks. It is difficult and time-consuming to construct such a degradation pool. While in our PDM, the blur kernels and noises are automatically learned via adversarial training, which is much easier.
	
	\vspace{0.02\linewidth}
	\noindent\textbf{Probabilistic degradation model.}
	To model the random factors in degradations,  in~\cite{gan_first} and~\cite{pseudosr}, the HR images are concatenated with random vectors before it is degraded by a neural network. Although their models consider the random factors, they do not well decouple the degradations with the image content, and do not provide an explicit formulation of the degradation. The relationship between the random vectors and degradations is unclear, and we also do not know what kinds of degradations the model learns. As a result, it is difficult to adjust the model settings according to different scenarios.  Instead, in our PDM, we parameterize the degradation with blur kernel $\mathbf{k}$ and random noise $\mathbf{n}$. The explicit formulation enables us to adjust the model flexibly, which will be further discussed in Sec~\ref{sec_pdm}.
	
	\section{Learning the Degradation Distribution}
	The degradation process in Eq~\ref{downsample} actually contains two linear steps~\cite{dan}:
	\begin{equation}
		\small
		\left\{
		\begin{aligned}
			&\mathbf{y}^{clean} = (\mathbf{x}\otimes \mathbf{k})\downarrow_{s}\\
			&\mathbf{y}= \mathbf{y}^{clean} + \mathbf{n}
		\end{aligned}
		\right.,
	\end{equation}
	where $\mathbf{y}^{clean}$ is the blurred,  downscaled,  and clean image without noise. Intuitively, the two steps are mutually independent, as the blur kernels are mainly dependent on the properties of the camera lens while the noises are mainly related to the properties of sensors. Thus, the degradation distribution can be modeled as:
	\begin{equation}
		p_D(\mathbf{D}) = p_{k, n}(\mathbf{k}, \mathbf{n}) = p_{k}(\mathbf{k})p_{n}(\mathbf{n}).
	\end{equation}
	In this way, the distributions of $\mathbf{k}$ and $\mathbf{n}$ can be independently modeled to represent the distribution of $\mathbf{D}$.
	
	\subsection{Kernel Module}\label{kernel_model}
	To model the distribution of  the blur kernel $\mathbf{k}$, we define a priori random variable $\mathbf{z}_k$ which is subject to multi-dimensional normal distribution. Then we use a generative module to learn the mapping from $\mathbf{z}_k$ to $\mathbf{k}$: 
	\begin{equation}
		\mathbf{k} = netK(\mathbf{z}_k) \quad  \mathbf{z}_k \sim \mathcal{N}(\mathbf{0}, \mathbf{1}), 
	\end{equation}
	where $netK$ is the generative module represented by a convolutional neural network. 
	
	Without loss of generality, we first consider the spatially variant blur kernel \ie, the blur kernel for each pixel of  $\mathbf{x}$ is different. In that case, we have
	\begin{equation}
		\mathbf{z}_k \in \mathcal{R}^{f_k\times h \times w} \quad \mathbf{k} \in \mathcal{R}^{(k\times k)\times h \times w}, 
	\end{equation}
	where $f_k$ is the dimension of the normal distribution $\mathbf{z}_k$,  $k$ is the size of the blur kernel, $h$ and $w$ are the height and width of  $\mathbf{x}$, respectively. We add a $Softmax$ layer~\cite{softmax} at the end to guarantee that all elements of $\mathbf{k}$ sum to one. Generally,  the sizes of the convolutional weights in $netK$ are set as $3\times3$, which indicates that the learned blur kernels are spatially correlated. Otherwise, if the spatial size of all convolutional weights is set as $1\times1$, then the blur kernel of each pixel is learned independently. 
	
	The spatial variances of blur kernels are usually caused by lens distortions, which mainly exit around the corners of the images. In most cases, the blur kernel could be approximated by a spatially invariant one,  which is a special case of the spatially variant blur kernel with $h=w=1$. And we have:
	\begin{equation}
		\mathbf{z}_k \in \mathcal{R}^{f_k\times 1 \times 1} \quad \mathbf{k} \in \mathcal{R}^{(k\times k)\times 1 \times 1}.
	\end{equation}
	The experiment results in Sec~\ref{compare} indicate that this approximation is already good enough for most datasets.
	
	\subsection{Noise Module}\label{noise_model}
	The second step of the degradation is adding noise to the blurred and downscaled image $\mathbf{y}^{clean}$. Most previous degradation models consider only AWGN (additive white gaussian noises), which are independent of the content of  $\mathbf{y}^{clean}$. In this case, the distribution of  $\mathbf{n}$ can also be expressed by a vanilla generative module:
	\begin{equation}
		\mathbf{n} = netN(\mathbf{z}_n) \quad  \mathbf{z}_n \sim \mathcal{N}(\mathbf{0}, \mathbf{1}), 
	\end{equation}
	where $netN$ is a convolutional neural network. We denote the height, width and number of channels of $\mathbf{y}$ as $h$, $w$ and $c$ respectively, then we have
	\begin{equation}
		\mathbf{z}_n \in \mathcal{R}^{f_n\times h \times w} \quad \mathbf{n} \in \mathcal{R}^{h\times w\times c},
	\end{equation}
	where $f_n$ is dimension of  the normal distribution $\mathbf{z}_n$. 
	
	In other methods~\cite{benchmarking_noise,photon_noise}, noise in the raw space $\mathbf{n}_{raw}$ is modeled as a combination of \textit{shot} and \textit{read} noise. And $\mathbf{n}_{raw}$ can be approximated by a heteroscedastic Gaussian distribution~\cite{unprocessing_raw}:
	\begin{equation}
		\mathbf{n}_{raw} \sim \mathcal{N}(\mathbf{0}, \sigma_{read}  + \sigma_{shot}  \mathbf{y}^{clean}), 
	\end{equation}
	where $\sigma_{read}$ and $\sigma_{shot}$ are determined by the camera sensor's analog and digital gains. Since $\mathbf{n}$ is originated from $\mathbf{n}_{raw}$, it indicates that $\mathbf{n}$ is also related to the image content, and the distribution of $\mathbf{n}$ should be expressed as by a conditional generative module:
	\begin{equation}
		\mathbf{n} = netN(\mathbf{z}_n, \mathbf{y}^{clean}) \quad  \mathbf{z}_n \sim \mathcal{N}(\mathbf{0}, \mathbf{1}).
	\end{equation}
	
	Accordingly, we can also adjust the sizes of convolutional weights in $netN$ to determine whether the noises are spatially correlated or not.
	
	\subsection{Probabilistic Degradation Model}~\label{sec_pdm}
	The kernel module and noise module discussed above together form our probabilistic degradation model (PDM), which is described in Figure~\ref{deg_model}. PDM can be used to synthesize HR-LR pairs: 
	\begin{equation}
		\small
		\left\{
		\begin{aligned}
			&\mathbf{y}^{clean}_{ref}  = 
			(\mathbf{x}_{ref}\otimes netK(\mathbf{z}_k))\downarrow_{s} \quad
			\mathbf{z}_k \sim \mathcal{N}(\mathbf{0}, \mathbf{1}) \\
			&\mathbf{y}_{ref} =
			\mathbf{y}^{clean}_{ref}+ netN(\mathbf{z}_n, \mathbf{y}^{clean}_{ref})  \quad \mathbf{z}_n \sim \mathcal{N}(\mathbf{0}, \mathbf{1})
		\end{aligned}
		\right.,
	\end{equation}
	where $\mathbf{x}_{ref}$ is the referring HR image and $\{\mathbf{x}_{ref}, \mathbf{y}_{ref}\}$ forms a paired training sample for the SR model. 
	
	Our PDM is optimized via adversarial training, which encourages $\mathbf{y}_{ref}$ to be similar with the test images $\mathbf{y}_{test}$. Additionally, we assume that the noise $\mathbf{n}$ has zero mean. Thus, besides the adversarial loss $l_{adv}$, we add an extra regularizer about noise $\mathbf{n}$: 
	\begin{equation}
		l_{reg} =  \|\mathbf{n}\|_2^2.
	\end{equation}
	Then the  total loss function of the degradation model is:
	\begin{equation}
		l_{total} = l_{adv} + \lambda 	l_{reg},  
	\end{equation}
	where $\lambda$ is the weight for the regularizer term. In all of our experiments, we set $\lambda=100$ to balance the magnitude of the two losses. 
	
	Compared with previous degradation models, PDM has three advantages:
	
	Firstly, PDM is able to model more diverse degradations. It allows one HR image to be degraded into multiple LR images. Thus, with the same number of HR images, PDM can generate more diverse LR images and provide the SR model with more training samples, which may better cover degradations of test images. As a result, PDM could bridge the gap between training and test datasets, and help the SR model to perform better on test images.
	
	Secondly, the prior knowledge about the degradations can be easily incorporated into PDM, which may encourage it to learn the degradations better. For example, if we observe that the blur is nearly uniform among a single image, then we can adjust the shape of $\mathbf{z}_k$ and $\mathbf{k}$ to learn only spatially invariant blur kernels. Such prior knowledge helps reduce the learning space of PDM and may encourage make it easier to be trained. 
	
	At last, PDM formulates the degradation process as a linear function, and the learned degradations can only impose a limited influence on the image content. In this way, it better decouples the degradations with image content and could focus on learning the degradations. In most previous methods, to ensure that $\mathbf{y}_{ref}$ has consistent content with $\mathbf{x}_{ref}$, it is usually guided by a bicubically downscaled version of $\mathbf{x}_{ref}$~\cite{gan_first,dasr}. However, this guidance may be inappropriate, especially when the test images are heavily blurred. Instead, in our PDM, due to the well-constrained blur kernels and noises, the content of  $\mathbf{y}_{ref}$ is inherently consistent with $\mathbf{x}_{ref}$. As a result, PDM could avoid the limitation of extra guidance and focus on learning the degradations.
	
	\subsection{A Unified Framework for Blind SR}
	In~\cite{fssr} and~\cite{dasr}, the training of the degradation model and the SR model are separate, \ie they firstly train a degradation model and then use the trained degradation model to help train the SR model. This two-step training method is time-consuming but is necessary for their method, because their highly nonlinear degradation models will produce undesirable results at the beginning of the training, which may mislead the optimization of the SR model. However, in our method, since PDM is better constrained and easier to be trained, it works well to train PDM and the SR model simultaneously. In this way, PDM can be integrated with any SR model to form a unified framework for blind SR, which is called PDM-SR (or PDM-SRGAN if adversarial loss and perceptual loss~\cite{srgan} are also adopted in the training of the SR model). 
	
	\section{Experiments}
	\subsection{Experimental Setup}
	\noindent\textbf{Datasets.}
	We conduct experiments mainly on five datasets, \ie, track2 of NTIRE2017~\cite{ntire2017},  track2  and track4 of NTIRE2018~\cite{ntire2018}, and track1 and track2 of NTIRE2020~\cite{ntire2020}. The details of different datasets are shown in Table~\ref{datasets}. The first three datasets provide $800$, $800$, and $3200$ paired HR-LR images for training and $100$ paired HR-LR images for validation.  But in this paper, we mainly study the cases where paired samples are not available. Thus, for each dataset, we use only the first half of HR images, and the second half of LR images for training.  For track1 and track2 of NTIRE2020,  since their provided training samples are already unpaired, we directly use all images for training.
	
	\begin{table}[h]
		\centering
		\setlength{\tabcolsep}{0.15cm}
		\caption{The details about different datasets. For datasets that have paired training samples, we use only the first half of HR images and the last half of LR images.}\label{datasets}
		\vspace{0.02\linewidth}
		\resizebox{\linewidth}{!}{
			\begin{tabular}{c|ccc|cc}
				\hline
				\multirow{2}{*}{Datasets} & \multicolumn{3}{c|}{Train} & \multicolumn{2}{c}{Validation} \\ \cline{2-6} 
				& LR (used)     & HR (used)     & paired  & LR             & HR            \\ \hline
				2017Track2~\cite{ntire2017}
				& $800(401\sim800)$   & $800(1\sim400)$   & \checkmark         & $100$          & $100$    \\
				2018Track2~\cite{ntire2018}
				& $800(401\sim800)$   & $800(1\sim400)$   & \checkmark       & $100$          & $100$      \\
				2018Track4~\cite{ntire2018}
				& $3200(1600\sim3200)$   & $3200(1\sim1600)$   & \checkmark         & $100$          & $100$     \\
				2020Track1~\cite{ntire2020}
				& $2650(1\sim2650)$  & $800(1~\sim800)$  & $\times$         & $100$          & $100$           \\
				2020Track2~\cite{ntire2020}
				& $2229(1\sim2229)$  & $800(1\sim800)$  &  $\times$          & $100$           & $0$           \\ \hline
		\end{tabular}}
	\end{table}
	
	\vspace{0.02\linewidth}
	\noindent\textbf{Metrics.}
	For validation sets that have ground-truth (GT) images, \ie, validation sets of 2017Track2, 2018Track2, 2018Track4 and 2020Track1, we use PSNR, SSIM~\cite{ssim}, and LPIPS~\cite{lpips} as evaluating metrics. The images are evaluated in RGB space and the LPIPS is calculated with the \textit{AlexNet}~\cite{alex}. Especially, for validation sets of 2018Track2 and 2018Track4, we use the officially provided scoring function\footnote{https://competitions.codalab.org/my/datasets/download/1ed1e509-86ed-4dc1-be3f-4bca1dbdbd83} to calculate PSNR and SSIM, in which the reference image is shifted around to find the best PSNR and SSIM.  For the validation set of 2020Track2, which has no GT images, we use the non-reference metric NIQE~\cite{niqe}.
	
	\vspace{0.02\linewidth}
	\noindent\textbf{Implement details.}
	We use different settings of PDM for different datasets. For all datasets, the settings of the kernel module are shared. The dimension of $\mathbf{z}_k$ is set as $f_{k} = 64$, and the size of the blur kernel is set as $21\times 21$. For simplicity, we assume that blur kernels in datasets are spatially invariant. For 2017Track2, since the test images are clean and have almost no noise, we omit the noise module in PDM. For the other three datasets, the dimension of $\mathbf{z}_n$ is set as $f_{n} = 3$, and the sizes of convolutional weights are set as $3\times3$. We use PatchGAN discriminator~\cite{cyclegan} as the discriminators for adversarial training. The SR model is chosen according to the compared methods. For fair comparisons, all compared methods share the same SR model. In this paper, there are two cases: the baseline EDSR~\cite{edsr} and RRDB~\cite{esrgan}.
	
	\vspace{0.02\linewidth}
	\noindent\textbf{Training.}
	We crop the HR images into $128\times128$ and the LR images into $32\times32$ for training. The batch size is set as $32$. All of our models are trained for $2\times10^5$ steps on a single RTX 2080Ti GPU. We use Adam~\cite{adam} as the optimizer. The learning rates for all models are set as $2\times10^{-4}$ at the beginning and decayed by half for each $5000$ steps.
	
	\begin{table*}[t]
		\centering
		\setlength{\tabcolsep}{0.2cm}
		\caption{The comparisons between different methods. The referring methods are implemented by ourselves. The SR models are set as baseline EDSR and the scale factors are $\times4$. `*' denotes that the metrics are calculated via different scripts. $\uparrow$ denotes the larger the better. $\downarrow$ denotes the smaller the better. The best results are denoted in {\color{red}red}, and the second best are denoted in {\color{blue}blue}.}\label{learned_deg_compare}
		\vspace{0.01\linewidth}
		\resizebox{\linewidth}{!}{
			\begin{tabular}{c|ccc|ccc|ccc|ccc}
				\hline
				\multirow{2}{*}{Methods} & \multicolumn{3}{c|}{2017Track2} & \multicolumn{3}{c|}{2018Track2} & \multicolumn{3}{c|}{2018Track4} & \multicolumn{3}{c}{2020Track1} \\ \cline{2-13} 
				& PSNR$\uparrow$     & SSIM$\uparrow$      & LPIPS$\downarrow$
				& *PSNR$\uparrow$     &*SSIM$\uparrow$      & LPIPS$\downarrow$
				& *PSNR$\uparrow$     &*SSIM$\uparrow$      & LPIPS$\downarrow$
				& PSNR$\uparrow$     & SSIM$\uparrow$      & LPIPS$\downarrow$    \\ \hline
				Bicubic
				&$21.73$  & $0.5731$ & $0.5430$
				& $20.58$ &$0.5304$ & $0.7929$
				&$20.26$ & $0.5101$ & $0.7970$
				& \color{blue}$25.51$ & $0.6731$ & $0.6414$\\
				EDSR~\cite{edsr}
				&$21.58$  & $0.5646$ & $0.4697$
				& $20.43$ &$0.5045$ & $0.7778$
				&$20.07$ & $0.4771$ & $0.7831$
				& $25.34$ & $0.6391$ & $0.6074$\\
				DSGAN-SR~\cite{fssr}
				&$20.18$  & $0.5165$ & $0.4314$
				& $20.67$ &\color{blue}$0.5829$ & $0.5293$
				&$20.26$ & $0.4402$ & $0.5381$
				& $23.29$ & $0.6631$ & $0.3295$\\
				CinCGAN~\cite{cingan} 
				& $19.04$  & $0.4451$  & $0.3847$
				& $20.10$  & $0.4631$  & \color{blue}$0.4748$
				& $20.09$  & $0.4680$  & \color{blue}$0.4903$
				& $21.70$  & $0.5814$ & \color{blue}$0.3386$ \\
				CycleSR~\cite{cyclesr}
				& $20.70$  & $0.5242$  & $0.4798$
				&\color{red} $21.36$  &$0.5291$  & $0.6390$
				& \color{blue}$20.65$  & $0.4980$  & $0.6574$
				&$25.48$  &\color{blue}$0.7259$ & $0.3641$ \\
				Maeda \etal~\cite{pseudosr}
				& $19.23$  & $0.4754$  & $0.3667$
				& $19.90$  & $0.4728$  & $0.4897$
				& $18.57$  & $0.4085$  & $0.5322$
				& $20.06$  & $0.5368$ & $0.4074$ \\
				Bulat \etal~\cite{gan_first}
				& $19.84$  & $0.5020$  & $0.4115$
				& $20.27$  & $0.4488$  & $0.6668$
				&\color{red}$20.91$  &\color{blue} $0.5408$  & $0.5918$
				& $21.49$  & $0.5553$ & $0.4935$ \\
				\hline
				PDM-SRGAN 
				&\color{blue}$23.43$  &\color{blue}$0.6412$ &\color{red}$0.2475$          
				&$20.32$  & $0.5257$  & \color{red}$0.4074$ 
				& $20.25$  & $0.5307$  &\color{red}$0.4415$ 
				& $24.56$  & $0.6630$ & \color{red}$0.2716$ \\
				PDM-SR 
				&\color{red}$23.69$  &\color{red} $0.6725$ &\color{blue}$0.3427$
				&\color{blue}$20.85$ &\color{red} $0.5870$&$0.5240$
				&$20.32$&\color{red} $0.5611$  &$0.5282$ 
				&\color{red}$26.80$  & \color{red}$0.7470$  & $0.3601$ \\
				\hline
			\end{tabular}
		}
	\end{table*}
	\begin{figure*}[h]
		\centering
		\includegraphics[width=\linewidth]{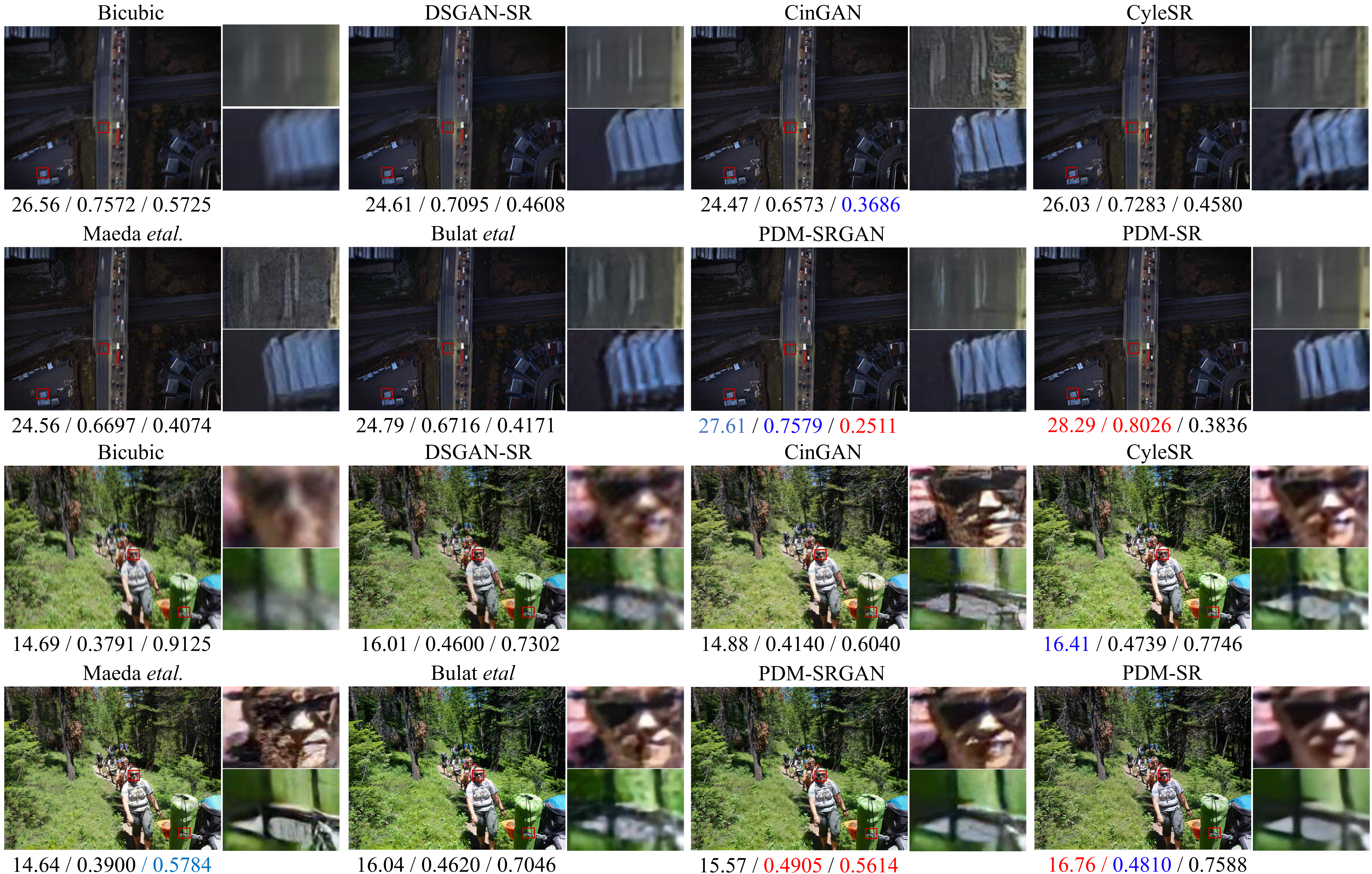}
		\caption{Visual comparisons on image \textit{0827x4} of 2017Track2 (top) and image \textit{0860x4m} of 2018Track2 (bottom). The \textit{PSNR / SSIM / NIQE} scores are denoted below the SR results respectively. Red: the best. Blue: the second best. \textbf{Zoom in for best view}.}\label{vis1}
		\vspace{-0.02\linewidth}
	\end{figure*}
	
	\subsection{Compared with Other Methods}\label{compare}
	\noindent\textbf{Compared with re-implemented methods.}
	As the codes or pretrained weights for most previous methods are not publicly available, we re-implement some of them for comparisons, including Bulat \etal~\cite{gan_first}, CinCGAN~\cite{cingan}, DSGAN-SR~\cite{fssr}, CycleSR~\cite{cyclesr}, and Maeda \etal~\cite{pseudosr}. To make fair comparisons, the peripheral settings, such as architectures of generators, discriminators, SR models \etc,  are kept the same. For example, the SR models of all referring methods are set as baseline EDSR~\cite{edsr}. All other details are set according to their published papers. The codes for implementations are included in the supplementary files. 
	
	As the referring methods include both PSNR-oriented (\ie, the SR model is supervised by L1/L2 loss) and perceptual-oriented (\ie, the SR model is supervised by perceptual loss~\cite{srgan,esrgan}) ones,  we also provide two versions of our methods, \ie, PDM-SR and PDM-SRGAN for comparisons.  The results of bicubic interpolation and baseline EDSR are also provided. As shown in Table~\ref{learned_deg_compare}, in terms of LPIPS, PDM-SRGAN performs much better than other methods. In terms of  PSNR and SSIM, PDM-SR also achieves the best overall performance. Especially on SSIM, PDM-SR outperforms all other methods by a large margin. 
	
	It is worth noting that our method has the greatest advantages on 2017Track2. It can be attributed to two reasons:
	\begin{itemize}
		\item  The LR images in 2017Track2 are blurred by various kernels, and the variety of its degradations may be larger than that in other datasets. PDM models the degradations as a distribution and is better at handling various degradations. 
		\item The test images of 2017Track2 are relatively clean. It is easy for us to incorporate this prior knowledge into PDM by removing its noise module. In this way, PDM can focus on learning the various blur kernels. While the degradation models of other methods are highly non-linear, which is difficult to be adjusted.
	\end{itemize}
	
	We also provide comparisons in Figure~\ref{vis1}. The LR image \textit{0827x4} from 2017Track2 is heavily blurred. Its SR result will likely have undesirable artifacts. As one can see, the results super-resolved by other methods are still blurry, while PDM-SR succeeds to remove the blur. The LR image \textit{0860x4m} in 2018Track2 suffers from complex noises. And as shown in the figure, the SR result of PDM-SR is relatively cleaner than that of other methods, which indicates that PDM can also better model the random noises.
	
	We need to note that although perceptual-oriented methods achieve relatively better LPIPS scores, their SR images may contain unrealistic distortions. It indicates that these metrics often fail to measure the visual quality of SR images~\cite{bsrgan}. A better metric is MOS (Mean Object Score)~\cite{lpips}, which, however,  is too expensive. As an alternative, we provide more visual comparisons in the supplementary files. 
	
	\begin{figure*}[t]
		\centering
		\includegraphics[width=\linewidth]{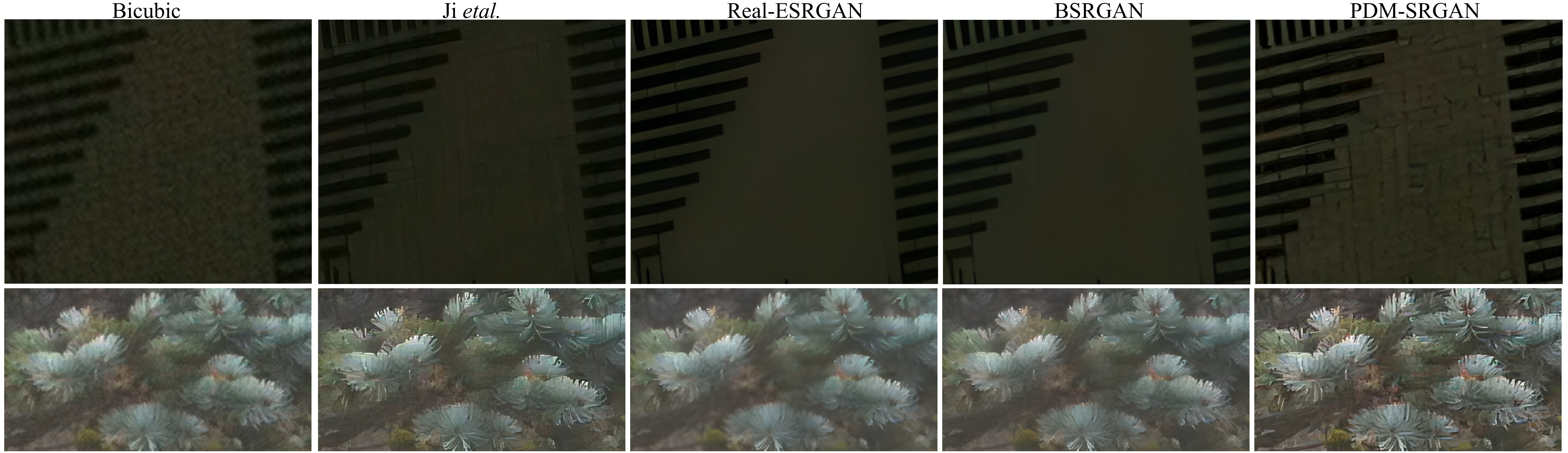}
		\caption{Visual comparisons on image \textit{0010} (top) and image \textit{0097} (bottom) of 2020Track2. \textbf{Zoom in for best view}.}\label{vis3}
		\vspace{-0.02\linewidth}
	\end{figure*}
	
\begin{figure}[t]
		\centering
		\includegraphics[width=\linewidth]{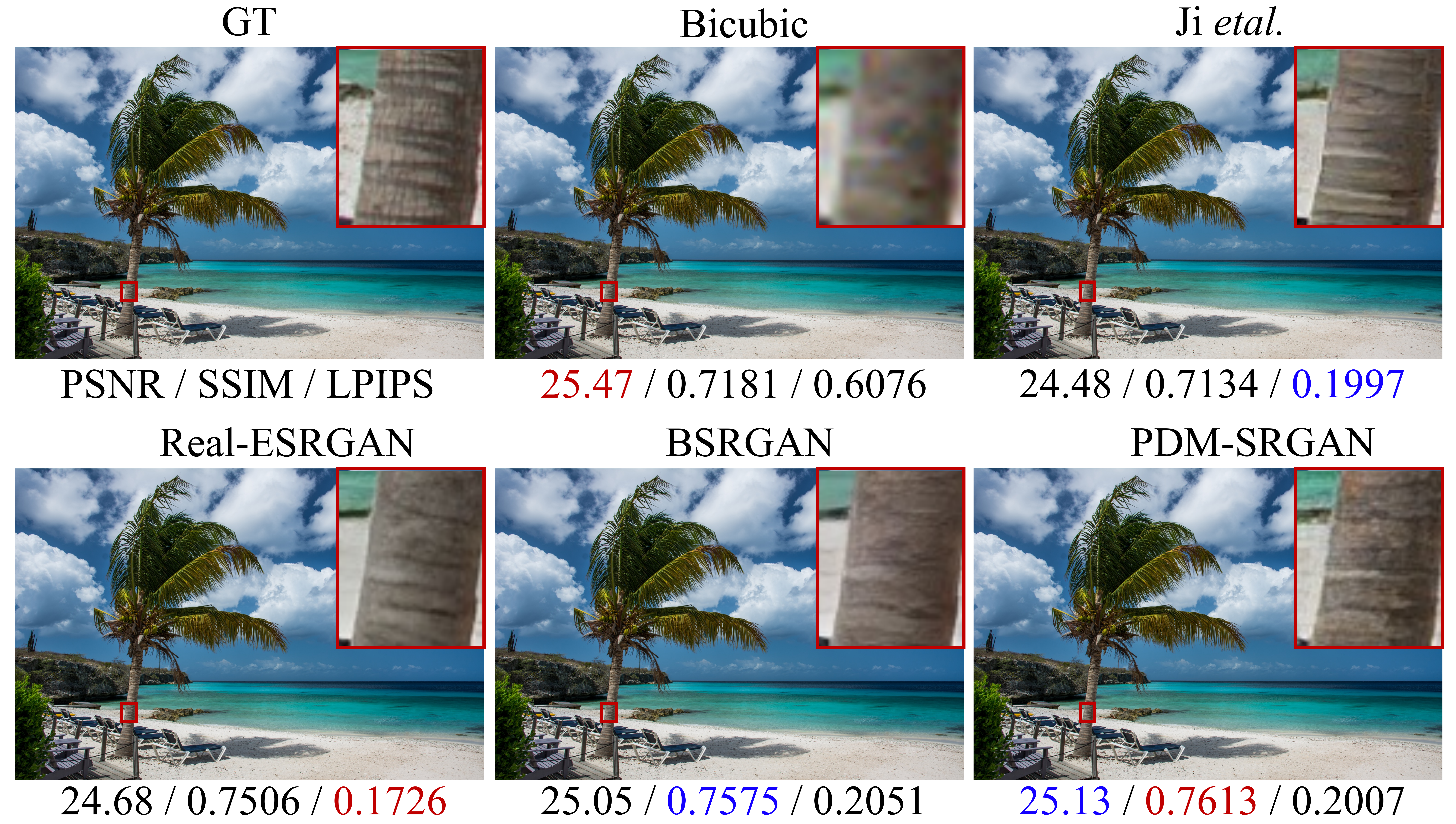}
		\caption{Visual comparisons on image \textit{0804} of 2020Track1. Red: the best. Blue: the second best. \textbf{Zoom in for best view}.}\label{vis2}
		\vspace{-0.03\linewidth}
	\end{figure}

	\vspace{0.02\linewidth}
	\noindent\textbf{Compared with pretrained methods.}
	Some methods provide the codes and pretrained weights,  including Ji \etal~\cite{ji2020real}, Real-ESRGAN~\cite{real-esrgan}, and BSRGAN~\cite{bsrgan}. We directly compare our method with their provided models. As the referring methods are perceptual-oriented, we provide the results of PDM-SRGAN for comparison. The SR models of the referring methods are RRDB~\cite{esrgan}, we also use the same SR model in our method. As shown in Table~\ref{pretrained_compare}, PDM-SRGAN achieves the best SSIM and LPIPS on 2020Track1 and the best NIQE on 2020Track2. We further provide visual comparisons in Figure~\ref{vis2} and Figure~\ref{vis3}. As one can see, the SR results produced by the referring methods are more likely to be over smooth. For example, in Figure~\ref{vis2}, the rough textures of the trunk are nearly lost in SR results of  Real-ESRGAN and BSRGAN, while these details are better preserved in PDM-SRGAN. And in the top row of Figure~\ref{vis3}, the cracks on the wall are also better preserved in PDM-SRGAN. This is because the other methods tend to consider excessive kinds of degradations and their SR models may tend to produce the average results for all cases. Instead, our PDM learns only the degradation distribution of the test images and helps train a specific SR model. Thus, PDM-SRGAN could produce SR results with more fine-grained details.
	\begin{table}[h]
		\centering
		\caption{Comparisons between different methods. The referring methods are tested with their officially provided models. The scale factor is $\times4$. $\uparrow$: the larger the better. $\downarrow$: the smaller the better. {\color{red} Red}: the best. {\color{blue} Blue}: the second best.}\label{pretrained_compare}
		\vspace{0.02\linewidth}
		\setlength{\tabcolsep}{0.2cm}
		\resizebox{\linewidth}{!}{
			\begin{tabular}{c|ccc|c}
				\hline
				\multirow{2}{*}{Methods} & \multicolumn{3}{c|}{2020Track1} & 2020Track2 \\ \cline{2-5} 
				& PSNR$\uparrow$     & SSIM$\uparrow$      & LPIPS$\downarrow$    & NIQE$\downarrow$       \\ \hline
				Ji \etal~\cite{ji2020real}
				&\color{red}$24.83$  & $0.6622$  &\color{blue} $0.2306$ & $4.12$     \\
				BSRGAN~\cite{bsrgan}
				& $24.55$  & $0.6688$  & $0.2690$ \color{blue}&\color{blue} $3.62$     \\
				Real-ESRGAN~\cite{real-esrgan}
				&\color{blue} $24.67$  &\color{blue} $0.6872$  & $0.2537$ & $3.75$     \\
				PDM-SRGAN
				&$24.43$  &\color{red}$0.7134$  &\color{red}$0.2243$ &\color{red}$3.19$     
				\\ \hline
			\end{tabular}
			\vspace{-0.05\linewidth}
		}
	\end{table}

	\subsection{Learned Degradations}
\begin{figure}[h]
		\centering
		\includegraphics[width=\linewidth]{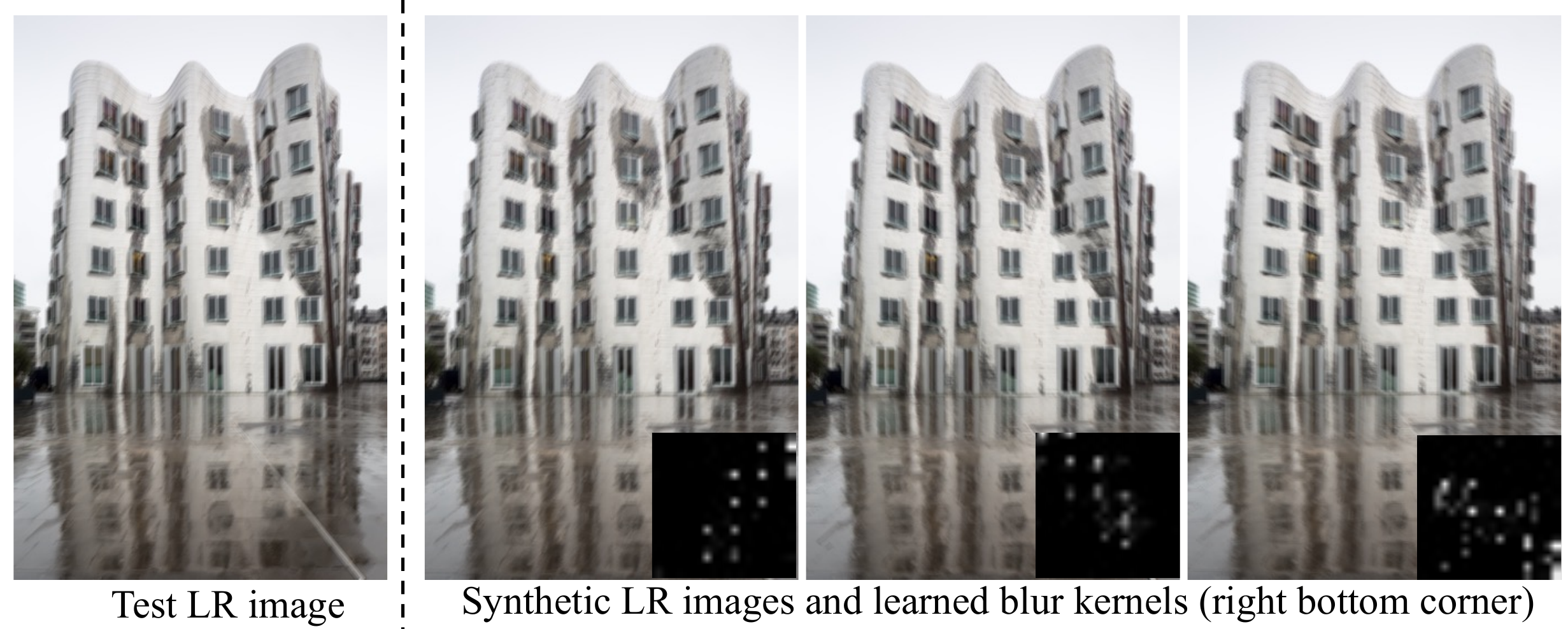}
		\caption{The generated LR images and blur kernels learned from 2017Track2.}\label{learned_kernel}
		\vspace{-0.05\linewidth}
	\end{figure}

	In this section, we study the degradations learned by our PDM. As shown in Figure~\ref{learned_kernel}, the blur kernels learned by PDM are quite different from gaussian ones. The support of learned kernels is scattered instead of being compact. This may be attributed to the downsampling operation:  the blur kernel is convolved with the HR image while the supervision is applied only to the LR image. The scattered blur kernels seem to be unreasonable, but the blur caused by these kernels may be non-distinguishable from that of real ones, and thus the training samples generated by PDM can help train a better SR model on this dataset.
	
\begin{figure}[h]
		\centering
		\includegraphics[width=\linewidth]{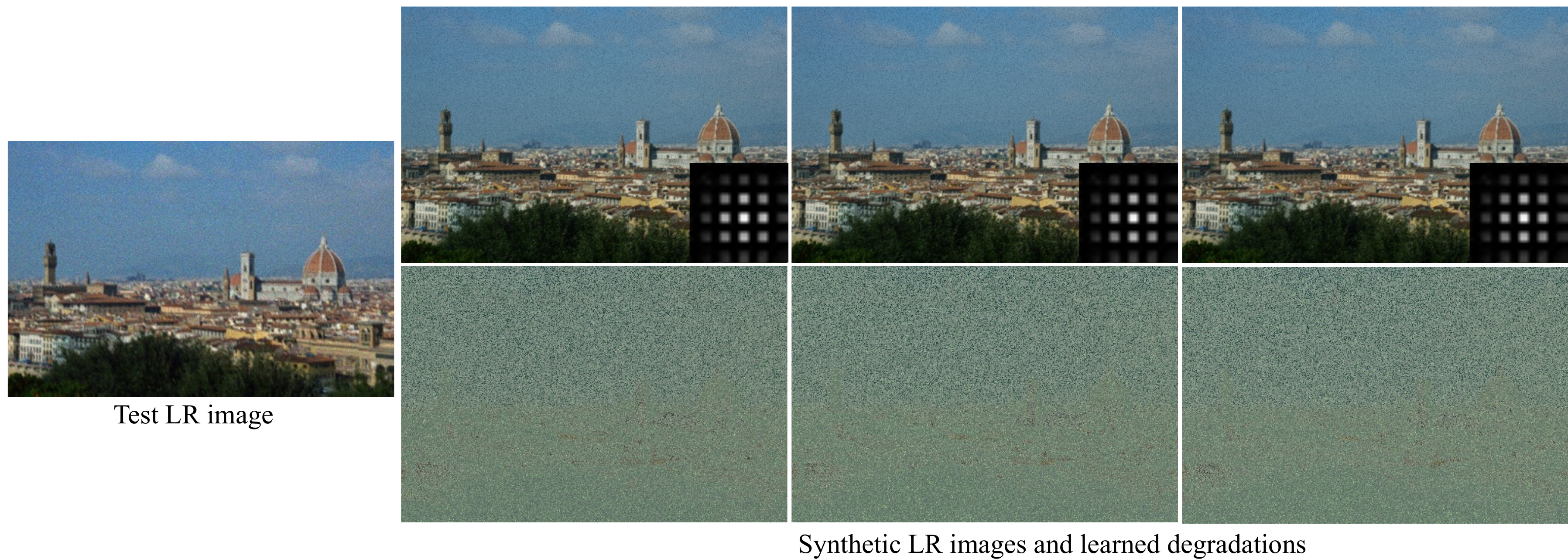}
		\caption{The generated LR images and degradations learned from 2018Track4. The blur kernel is plotted at the right bottom corner of the LR image, and the corresponding noise is plotted below.}\label{learned_noise}
		\vspace{-0.03\linewidth}
	\end{figure}

	As shown in Figure~\ref{learned_noise}, the blur kernels learned from 2018Track4 are different from that of 2017Track2: the blur kernels from 2017Track2 are asymmetric and various while those from 2018Track4 are symmetric and similar to each other. The learned noise is colorful, whose color is related to the content of the image.
	
	
	\subsection{Ablation Studies}
	\noindent \textbf{Probabilistic v.s. determined.} To better demonstrate the superiority of our PDM over deterministic degradation models,  we perform comparative experiments by substituting $\mathbf{z}_k$ and $\mathbf{z}_n$ in PDM to fixed zero inputs. The experiments are done on 2018Track4 because the degradations involved in it are of a larger variety. As shown in Table~\ref{aba_disritbution}, if either $\mathbf{z}_k$ or $\mathbf{z}_n$ is randomly sampled, which means that PDM can learn the distribution of blur kernels or noises, the SR results are better than both $\mathbf{z}_k$ and $\mathbf{z}_n$ are fixed as zeros. If both $\mathbf{z}_k$ and $\mathbf{z}_n$ are randomly sampled, the SR results are the best. It indicates that the probabilistic degradation model can help them achieve better SR performance. 
	
	\begin{table}[h]
		\centering
		\caption{Ablation studies about learning the degradation distributions. Results are reported on 2018Track4.}\label{aba_disritbution}
		\vspace{0.02\linewidth}
		\setlength{\tabcolsep}{0.5cm}
		\resizebox{\linewidth}{!}{
			\begin{tabular}{cc|ccc}
				\hline
				$\mathbf{z}_k$        & $\mathbf{z}_n$
				& PSNR$\uparrow$    & SSIM$\uparrow$    & LPIPS$\downarrow$    \\ \hline
				&           
				& $19.27$ & $0.4629$ & $0.6021$ \\
				\checkmark &           
				& $20.12$ & $0.5231$ & $0.5683$ \\
				& \checkmark 
				& $20.18$ & $0.5474$ & $0.5468$ \\
				\checkmark & \checkmark 
				& $\mathbf{20.32}$ &$\mathbf{0.5611}$ & $\mathbf{0.5282}$ \\ \hline
		\end{tabular}}
		\vspace{-0.03\linewidth}
	\end{table}
	\vspace{0.02\linewidth}
	\noindent \textbf{Content-independent v.s. content-dependent.} As we have discussed in Sec~\ref{kernel_model}, we assume that the blur kernels are independent of the content of the image, the blur kernels are generated only from randomly sampled vectors, \ie, $\mathbf{k} = netK(\mathbf{z}_k)$. To validate this assumption, two other experiments are performed: A) the blur kernels are completely dependent on the original HR image $\mathbf{x}$, \ie, $\mathbf{k} = netK(\mathbf{x})$; B) the blur kernels are partially dependent on $\mathbf{x}$, \ie, $\mathbf{k} = netK(\mathbf{x}, \mathbf{z}_k)$. The experiments are performed on 2017Track2, whose LR images are relatively clean which can exclude the influence of random noises. As shown in Table~\ref{aba_dependent_k}, the SR results of the two experiments are worse than our method. It indicates that content-independent blur kernel is a better assumption for 2017Track2.
	
	\begin{table}[h]
		\centering
		\caption{Exploring the dependences between the blur kernels and the content of images. Results are reported on 2017Track2.}\label{aba_dependent_k}
		\vspace{0.02\linewidth}
		\setlength{\tabcolsep}{0.5cm}
		\resizebox{\linewidth}{!}{
			\begin{tabular}{c|ccc}
				\hline
				Experiments & PSNR$\uparrow$    & SSIM$\uparrow$    & LPIPS$\downarrow$    \\ \hline
				$\mathbf{k} = netK(\mathbf{x})$    
				& $19.39$ & $0.5117$ & $0.3485$ \\
				$\mathbf{k} = netK(\mathbf{x}, \mathbf{z}_k)$         
				& $20.19$ & $0.5435$ & $03454$ \\
				$\mathbf{k} = netK(\mathbf{z}_k)$
				& $\mathbf{23.69}$ & $\mathbf{0.6725}$ & $\mathbf{0.3427}$ \\ 
				\hline
		\end{tabular}}
	\end{table}
	
	Similarly, we also explore the dependences between the noise and the content of the image. In our method, the noise $\mathbf{n}$ is partially dependent on $\mathbf{y}^{clean}_{ref}$, \ie, $\mathbf{n} = netN(\mathbf{y}^{clean}_{ref }, \mathbf{z}_n)$. For comparisons, two other experiments are performed: A) the noises are completely dependent on $\mathbf{y}^{clean}_{ref}$, \ie, $\mathbf{n} = netN(\mathbf{y}^{clean}_{ref})$; B) the noises are completely independent of $\mathbf{y}^{clean}_{ref}$, \ie,  $\mathbf{n} = netN(\mathbf{z}_n)$. The experiments are performed on 2020Track1, whose LR images are less blurry. As shown in Table~\ref{aba_dependent_n}, when $\mathbf{n}$ is partially dependent on $\mathbf{y}^{clean}_{ref}$, the method gets the best overall performance. 
	
	We need to note that these assumptions may be violated in other datasets, and the model settings can be easily adjusted according to different prior knowledge. 
	
	\begin{table}[h]
		\centering
		\caption{Exploring the dependences between the random noises and the content of images. Results are reported on 2020Track1.}\label{aba_dependent_n}
		\vspace{0.02\linewidth}
		\setlength{\tabcolsep}{0.5cm}
		\resizebox{\linewidth}{!}{
			\begin{tabular}{c|ccc}
				\hline
				Experiments & PSNR$\uparrow$    & SSIM$\uparrow$    & LPIPS$\downarrow$    \\ \hline
				$\mathbf{n} = netN(\mathbf{z}_n)$    
				& $26.14$ & $0.7352$ & $0.3641$ \\
				$\mathbf{n} = netN(\mathbf{y}^{clean}_{ref})$         
				& $25.81$ & $0.7274$ & $\mathbf{0.3432}$ \\
				$\mathbf{n} = netN(\mathbf{y}^{clean}_{ref}, \mathbf{z}_n)$
				& $\mathbf{26.80}$ & $\mathbf{0.7470}$ & $0.3601$ \\ 
				\hline
		\end{tabular}}
		\vspace{-0.03\linewidth}
	\end{table}
	
	\section{Limitations and Discussions}
	The first limitation of our method is that PDM needs a certain amount of LR images to learn their degradations. If there is only one LR image, or LR images are taken from completely different sources,  it will be difficult for PDM to learn the degradation distribution. This is also a limitation for most existing degradation-learnings-based methods. In our method, this problem may be alleviated by using a powerful pretrained SR model, such as BSRGAN~\cite{bsrgan} or Real-ESRGAN~\cite{real-esrgan}. These pretrained models could serve as priors and reduce the demand for training data. 
	
	The second limitation is that the current version of PDM does not consider the corruptions introduced by JPEG compression, which will also influence the performance of SR. In future work, we may add an extra learnable JPEG compression module~\cite{jpeg} in PDM, which could enable PDM to simulate the JPEG corruptions.
	
	\section{Conclusion and Future Work}
	In this paper, we study the degradation function as a random variable and model its distribution as the joint distribution of the blur kernel $\mathbf{k}$ and random noise $\mathbf{n}$. In this way, the proposed probabilistic degradation model (PDM)  could better decouple the degradations with the content of the image and model the random factors in degradations. Compared with previous degradation models, PDM could generate HR-LR training samples with a larger variety of degradations, which may better cover the degradations of test LR images and help improve the performance of SR models on the test images. Moreover, the proposed PDM provides a flexible formulation of degradation, which can be easily adjusted according to different real scenarios. In the future, we may add an extra learnable JPEG compression module in PDM to further enable it to simulate the JPEG corruptions.
	
	\section*{Acknowledgement}
	This work was jointly supported by National Key Research and Development Program of China Grant No. 2018AAA0100400, National Natural Science Foundation of China (61721004, U1803261, and 61976132), Beijing Nova Program (Z201100006820079), Key Research Program of Frontier Sciences CAS Grant No. ZDBS-LY-JSC032, and CAS-AIR.
	{\small
		\bibliographystyle{ieee_fullname}
		\bibliography{egbib}
	}
	
\end{document}